\documentclass{article}
\usepackage[T1]{fontenc}
\usepackage{graphics}
\usepackage[latin1]{inputenc}

\makeatletter

\newcommand{\LyX}{L\kern-.1667em\lower.25em\hbox{Y}\kern-.125emX\spacefactor1000}

\makeatother

\begin{document}

\begin{center}
\textbf{\textsc{Open clusters or their remnants: B and V photometry of NGC\,1901 and NGC\,1252}} 
\end{center}

\begin{center}
D.B. Pavani\protect\( ^{1}\protect \), E. Bica\protect\( ^{1}\protect \), C.M. Dutra\protect\( ^{1}\protect \), H. Dottori\protect\( ^{1}\protect \),
B.X. Santiago\protect\( ^{1}\protect \),G. Carranza\protect\( ^{2}\protect \), R.J. Díaz\protect\( ^{2}\protect \)
\end{center}

\begin{center}
\( ^{1} \)Universidade Federal do Rio Grande do Sul, IF, CP 15051, Porto Alegre
91501--970, RS, Brazil

\( ^{2} \)Observatorio Astronómico de Córdoba, Laprida 854, 5000, Córdoba,
Argentin
\end{center}

\begin{abstract}
Photometry in the B and V bands is presented for the southern stellar groups  NGC\,1901 and NGC\,1252. NGC\,1901 is often described as an open cluster while NGC 1252 consists of a concentration of about 20 stars centered ${\approx}$ 20$^{\prime}$ north of the original New General Catalogue coordinates, and at the southwest edge of the large region previously assigned to this object in the literature. NGC\,1901 has a clear main sequence and shares similarities with the Hyades. We derive a reddening value $E(B-V) = 0.04$, a distance from the Sun $d_{\odot} = 0.45$ kpc ($Z = -0.23$ kpc) and an age 0.6 ${\pm}$ 0.1 Gyr. NGC\,1901 is conclusively a physical system, dynamically comparable to or more evolved than the Hyades. The colour-magnitude diagram of NGC\,1252 suggests a turnoff and main sequence, and a total of 12 probable members. We simulated the Galactic field colour - magnitude diagram in the same direction and found it to be a poor match to NGC\,1252, suggesting  that NGC\,1252 is not a field fluctuation. Isochrone fitting to the probable members is consistent with $E(B-V) = 0.02$, $d_{\odot} = 0.64$ kpc ($Z = -0.46$ kpc) and an age 3 ${\pm}$ 1 Gyr.  NGC\,1252 cannot be ruled out as a physical group with the available data. If so, evidence is found that it is not a classical open cluster, but rather an open  cluster remnant.
\end{abstract}

\begin{center}
\end{center}

\section{Introduction}

Open clusters dynamically evolve and eventually dissolve. They appear to behave as weakly bound or slightly unbound systems. Catalogues include many poorly populated and/or low density stellar concentrations which may have different natures such as cluster remnants, parts of associations, quasi-associations and multiple star systems, or non-physical effects such as field fluctuations. Lod\'en (1977) presented a list of relatively loose young concentrations that probably have never been bound.  The Basel objects (e.g. Hassan 1974) are concentrations which appear to be physical stellar groups, in contrast to concentrations produced by random fluctuations in fields of Milky Way stars (e.g. Ruprecht\,46 - Carraro \& Patat 1995).

Wielen (1971) studied the evolution and dissolution time scale of open clusters. Recently, numerical simulations suggested that many concentrations may be open cluster remnants (e.g. de la Fuente Marcos 1998). Bassino et al. (2000) find evidence of a possible remnant - M\,73 (NGC 6994), but Carraro (2000) does not favour it. Bica et al. (2001) presented 34 dissolving star cluster candidates which are located at relatively high galactic latitudes ($|b| >$ 15$^{\circ}$) and are underpopulated with respect to usual open clusters, but they possess a significant number density contrast as compared to the Galactic field. The distinction between an open cluster and a remnant is not obvious both observationally and theoretically. An open  cluster remnant can be defined as a poorly populated concentration of stars as a result of the dynamical evolution of an initially more massive physical system. A widely accepted open cluster such as the Hyades appears to have lost 90 \% of the original stellar content (Weidemann et al. 1992). At the other extreme a binary star, the ultimate remains of a dissolved open cluster, is a cluster remnant. In the present study a cluster remnant is defined as a poorly populated physical concentration of stars  with enough members to show evolutionary sequences in a colour-magnitude diagram (CMD). 

NGC 1252 was discussed in Bica et al. (2001), who refer to a stellar concentration at the southwest edge of Bouchet \& Th\'e`s (1983) definition  of NGC 1252, which is a sparse  collection of stars in a much larger field. NGC\,1252 is described in the NGC as a star cluster containing 18 or 20 stars. It is located in Horologium with original coordinates (available e.g. in Sulentic \& Tifft 1973) corresponding to J2000 ${\alpha = 3 ^h10 ^m 31 ^s }$ , ${\delta = -58 ^{\circ} 08 ^{\prime} 22''}$ ($l = 274.^{\circ}59,\, b = -50.^{\circ}64$). All $\alpha$, $\delta$ values in the present study will refer to the epoch J2000.0. Bouchet \& Th\'e (1983, hereafter BT) observed 38 stars with UBVRI photometry in a 1$^{\circ}$ diameter region centered on the bright carbon star TW Horologii (${\alpha = 3 ^h12 ^m 33.2 ^s }$ , ${\delta = -57 ^{\circ} 19 ^{\prime} 18''}$ and $l = 273.^{\circ}30,\, b = -50.^{\circ}90$). They considered 16 as probable members, including TW Hor (BT\,38) itself, and estimated a distance from the Sun d$_{\odot}$ = 470 pc. Using CZC proper motions  Eggen (1984) concluded that BT$^{\prime}$s cluster was non-existent and that TW Hor (d$_{\odot}$ = 400 pc) probably belonged to the Hyades Supercluster. Using proper motions contained in the ACT and Hipparcos catalogues Baumgardt (1998) also concluded that the cluster as interpreted by BT is non-existent.

CCD photometry is obtained for NGC\,1252 as in Bica et al. (2001) at ${\alpha} = 3^h 10^m49^s, {\delta} = -57^{\circ} 46^{\prime}00''$ ($l = 274.^{\circ}08$, $b = -50.^{\circ}83)$. This concentration can be seen in Fig. 1 of BT as the small group encompassed by the bright stars BT\,1 and 13. An XDSS (second generation Digitized Sky Survey) blow-up of the concentration was shown in Fig. 4 of Bica et al. (2001). We emphasize that the present set of stars related to the concentration is not BT$^{\prime}$s set of members.

We also address the crucial point of the distinction between an \textit{open cluster}  and an \textit{open cluster remnant}. We deepen the photometry of the poorly populated open cluster or stellar group NGC 1901 for comparisons with NGC\,1252. NGC 1901  at ${\alpha}$ = ${5^h 18^m 11^s }$, ${\delta = -68{^\circ} 27^{\prime} 00''}$ ($ l = 279.^{\circ} 03$, $b = -33.^{\circ} 60$) is located in Dorado and was reported by Bok \& Bok (1960) as a loose stellar grouping. Sanduleak \& Philip (1968, hereafter SP) by means of BV photometry concluded that NGC\,1901 was a nearby Galactic stellar group projected onto the LMC. They found a turnoff near A0, a reddening value E(B-V)= 0.065 and  distance from the Sun d $_{\odot}$ = 330 pc. Murray et al. (1969) carried out astrometry for SP $^{\prime}$s stars with Cape Astrographic plates taken over more than 60 years. They  found that 14 of them have proper motions in common and called the object a star cluster. However, despite the conclusion of it being a physical system there remains the  doubt whether NGC\,1901 is  a classical open cluster or a remnant, or some transition state. In Bica et al. (2001) NGC\,1901 was  taken as a  comparison loose cluster and the number density contrast was important. Sect. 2 describes observations and reductions. Sect. 3 discusses the comparison object NGC\,1901. Sect. 4 discusses the properties of NGC\,1252. In Sect. 5 we compare the present objects with the Hyades and discuss the possibility of remnant physical systems. Concluding remarks are given in Sect. 6.

\section{Observations and reductions}

The observations of NGC\,1252 and NGC\,1901 were acquired at the C\'ordoba University Bosque Alegre Astrophysical Station, Argentina, on the night 28 December 1998, using a 30 cm Schmidt-Cassegrain telescope of the Universidade Federal do Rio Grande do Sul, temporarily installed at the site. CCD images with the B and V filters were collected with a KODAK KAF-0400 chip of 768${\times}$512 pixels and  size 9${\mu}m{\times}9{\mu}$m, corresponding to a field 12.$^{\prime}$4${\times}$8.$^{\prime}$3 on the sky. We employed 3 B and 2 V frames of exposure time 20 sec., and 1 B frame and 1 V frame of 40 sec. We illustrate a CCD V image of NGC\,1252 in Fig. 1 which includes most of the concentration. 

The reductions were carried out with IRAF starting with bias, dark, and flat field corrections. We combined the images for  better S/N ratios. DAOPHOT (Stetson 1992) was used to extract stars and derive instrumental magnitudes. We show in Fig. 2 the DAOPHOT (B-V) and V internal errors as a function of V mag for NGC\,1252: ${\epsilon}_{V}$ and ${\epsilon}_{(B-V)}$ attain 0.05 for V ${\approx}$ 15. We calibrated the data using stars from BT$^{\prime}$s photoeletric observations in the frame, since our main objective was to deepen their photometry in the stellar concentration area and check whether a significant MS could be present. All stars share the airmass X ${\approx}$ 1.19, and previous to the standard system transformation we corrected them for this effect. We used the site average atmospheric extinction coefficients $ K_V = 0.16$ and $K_{(B-V)} = 0.13$ (Minniti et al. 1989). The adopted standard stars are BT\,12, 14, 15, 16, 17, 18, 19 and 27. The resulting transformation equations to the standard system are

\begin{figure}
\resizebox{\hsize}{!}{\includegraphics{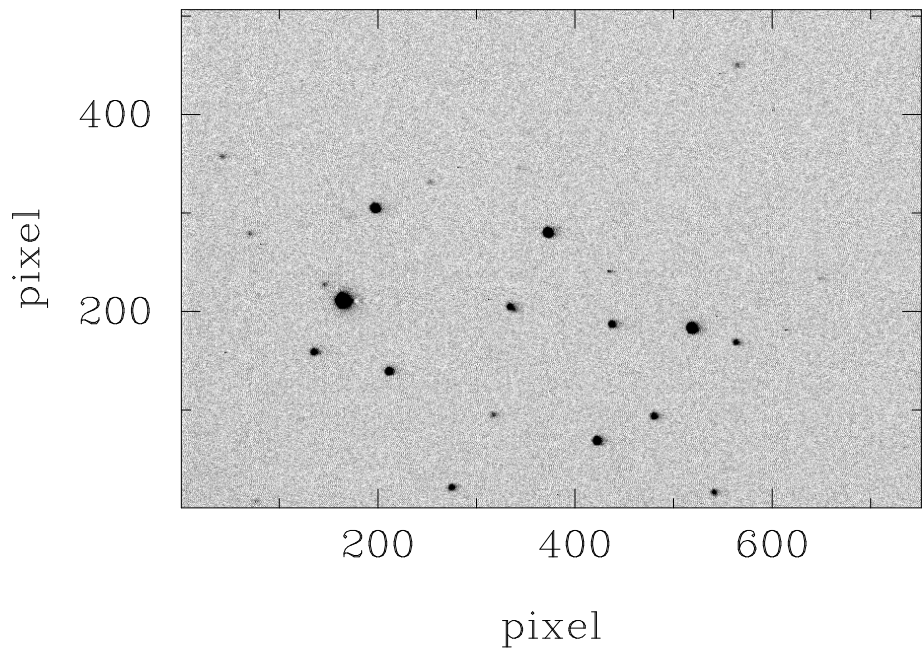}}
\caption[]{V  CCD image of NGC\,1252 with exposure 40 sec. Field is 12.$^{\prime}$4${\times}$8.$^{\prime}$3. Northwest is to the left and northeast to the top.}
\label{fig1}
\end{figure}

\hspace{1.5cm}

$V = 17.83 + v + 1.61{\times}(b-v)$

\hspace{1.5cm}

$(B-V) = -1.16 + 1.60{\times}(b-v)$

\hspace{1.5cm}

where $v$ and $(b-v)$ are the instrumental values corrected for atmospheric extinction. The external errors were estimated as the r.m.s. of the differences betwen standard and calculated values, and we obtained ${\rho}_V = 0.13$ and ${\rho}_{(B-V)} = 0.04$.

The bright stars BT\,1 (${\alpha} = {03 ^h10 ^m50.0 ^s}$, ${\delta} = {-57 ^{\circ} 42 ^{\prime}}$ ${ 06''}$) and BT\,11 (${\alpha} = {03 ^h11 ^m 09.0 ^s}$, ${\delta} = {-57 ^{\circ} 47 ^{\prime} 38''}$) were saturated in our frames. We adopt BT$^{\prime}$s photoeletric values for BT\,1 ($V = 8.68, (B-V) = 1.03$) and BT\,11 ($V = 10.50, (B-V) = 1.05$). For the bright star BT\,13 (${\alpha} = {03 ^h 10 ^m 39.2 ^s}$, ${\delta} = {-57^{\circ} 48 ^{\prime} 35.3''}$), located in the concentration just outside the CCD frames, we adopt BT$^{\prime}$s values V = 6.62, (B-V) = 0.89. The same applies to BT\,28 at ${\alpha} = {03 ^h10 ^m38.5 ^s}$, ${\delta} = {-57 ^{\circ} 47 ^{\prime} 20"}$ with $V = 11.95$ and $(B-V) = 0.47$.

We obtained CCD photometry for 12 stars in the NGC 1252 concentration. Among them 8 are also present in BT while 4 are newly observed. The results are shown in Table 1, by Col(s).: (1) identification, (2) and (3) J2000 equatorial coordinates, (4) and (5) stars with photoeletric V and (B-V) values from BT, (6) and (7) present CCD photometry in V and (B-V), and (8) membership.

\begin{table*}
\caption[]{CCD Photometric Results for NGC\,1252}
\label{tab1}
\renewcommand{\tabcolsep}{1.5mm}
\renewcommand{\arraystretch}{0.5}
\begin{tabular}{lccccccc}
\hline\hline
 Name$^{\mathrm{a}}$ & $\alpha$ & $\delta$& ~V ~&(B-V) & ~V ~&(B-V) & Membership$^{\mathrm{b}}$ \\
& h:m:s &$^{\circ}$:$^{\prime}$~~:$^{\prime\prime}$ & BT & BT & CCD & CCD &\\\hline
BT\,15 & 03:11:10.7 & -57:44:38 & 11.47 & 1.21 & 11.25 & 1.18 & lpm \\
BT\,12 & 03:10:50.2 & -57:47:11 & 11.95 & 0.50 & 11.67 & 0.55 & pm \\
BT\,17 & 03:11:02.6 & -57:41:48 & 11.92 & 1.51 & 11.99 & 1.54 & nm\\
BT\,18 & 03:10:45.0 & -57:43:23 & 12.49 & 0.73 & 12.40 & 0.70 & pm\\
BT\,14 & 03:11:04.4 & -57:46:22 & 12.91 & 0.60 & 12.93 & 0.63 & pm \\
BT\,27 & 03:10:56.5 & -57:47:48 & 13.17 & 0.49 & 13.22 & 0.55 & pm\\
BT\,16 & 03:10:59.8 & -57:44:43 & 13.21 & 0.78 & 13.40 & 0.78 & pm\\
BT\,19 & 03:10:42.4 & -57:42:06 & 13.21 & 0.83 & 13.49 & 0.73 & pm\\
GSC\,0849800928 & 03:10:51.6 & -57:49:21 & - & - & 14.63 & 0.95 & pm\\
GSC\,0849800945 & 03:11:10.2 & -57:48:26 & - & - & 14.67 & 1.10 & pm\\
GSC\,0849801321 & 03:10:51.7 & -57:40:06 & - & - & 15.29 & 0.35 & nm\\
GSC\,0849801024 & 03:10:46.5 & -57:45:20 & - & - & 15.45 & 0.75 & pm\\

\hline\hline

\end{tabular}
\begin{list}{}
\item {$^{\mathrm{a}}$\textit{BT} is from Bouchet \& Th\'e (1983), \textit{GSC} Guide Star Catalogue; $^{\mathrm{b}}$\textit{pm} means probable, \textit{lpm} less probable and \textit{nm} non-member.} 
\end{list}
\end{table*}

\begin{figure}
\resizebox{\hsize}{!}{\includegraphics{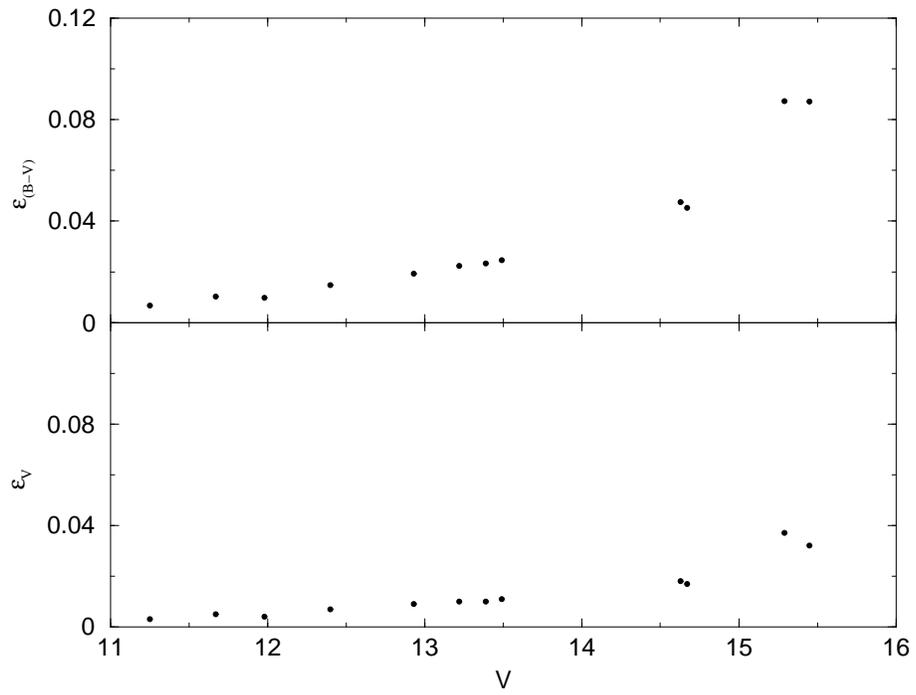}}
\caption[]{NGC\,1252: (B-V) internal errors (upper panel) and V internal errors (lower panel) as a function of V magnitude.}
\label{fig1}
\end{figure}

For NGC\,1901 we used 1 B and 1 V  10 sec. frames, and 3 B and 3 V 30 sec. frames. The object was observed at X ${\approx}$ 1.45, and the instrumental magnitudes were corrected accordingly. As standard stars we used SP\,7, 8, 9, 10, 12, 13, 14, 15 and 16. The distribution of the photometric internal errors is similar to that of NGC\,1252 in Fig. 2, except that ${\epsilon}_{(B-V)}$ attains 0.05 for V ${\approx}$ 14.5. The external errors are ${\rho}_V = 0.04$ and ${\rho}_{(B-V)} = 0.01$. The results are in Table 2 where we also indicate the SP values for comparison purposes. The columns follow those of Table 1, except two additional ones for Tycho proper motions (Sect. 3).

\section{The stellar group NGC\,1901}

We observed NGC\,1901$^{\prime}$s central concentration which can be seen in the chart provided by the WEBDA database (Mermilliod 1996) in the web interface http://obswww .unige.ch/webda. It has a diameter of ${\approx}$ 19$^{\prime}$. There occurs an extension 20$^{\prime}$ to the NW (Sanduleak \& Philip 1968), which presents proper motions comparable to those of the central stars (Murray et al. 1969). The V ${\times}$ (B-V) CMD derived from the CCD photometry (30 stars, Table 2) is shown in Fig. 3 as the large (previous stars) and small solid circles (photometry for the first time). A  gain of ${\Delta}$V ${\approx}$ 4 mag is obtained with respect to SP$^{\prime}$s photometry. The bright stars in the CCD area define a MS. The evolved star HD\,35294 (TYC2 573) was not included by SP in their sample, and it was saturated in our frames. We adopted Johnson V = 8.40, (B-V) = 0.70  values derived from the Tycho photometry, which are accurate for such bright stars. We superimposed on the CMD the NW extension stars (upwards open triangles) using SP values, which basically follow the upper MS and turnoff (TO) of the central stars. The SP stars to the south outside the CCD area (downwards open triangles) are located along the MS. Both the NW extension and the south stars are  compatible in the CMD with the bulk of central stars of NGC\,1901.

NGC\,1901 is projected onto the young LMC disc, and luminous LMC stars are expected. The LMC intermediate age disc (Bica et al. 1998, 1999) provides luminous AGB stars at the limit of the present photometry. The LMC RGB tip is below our limit, and the clump of red supergiants at an age ${\approx}$ 10 Myr, such as that in the LMC cluster NGC\,2004 (Bencivenni et al. 1991), is far too red (V ${\approx}$ 13.2, (B-V) ${\approx}$ 1.55) to contaminate the MS of NGC\,1901. We conclude that LMC intermediate colour supergiants and some bright AGB stars are the main field contributors. This can be seen in the Shapley III fields (Dolphin \& Hunter 1998) in the young LMC disc north of the bar, likewise the LMC part onto which NGC\,1901 is projected. Probable LMC members (Table 2) are located bluewards of the dashed line in Fig. 3. We cannot rule out some LMC contamination in the observed lower MS of NGC\,1901.

We employed solar metallicity Padova isochrones (Girardi et al. 2000) of different ages for best fit solutions (Fig. 3) and parameters. We find a reddening value $ E(B-V) = 0.04$ and an apparent distance modulus $(m-M) = 8.37$. The isochrones correspond to ages 0.5, 0.6 and 0.9 Gyr, from which we estimate an age 0.6 ${\pm}$ 0.1 Gyr for NGC\,1901, thus similar to that of the Hyades (WEBDA, Weidemann et al. 1998). The 0.6 Gyr isochrone basically fits a TO distribution and HD\,35294, which is thus a photometrically probable member. The latter star together with all those compatible with the 0.6 Gyr isochrone within total colour errors $\sqrt{\epsilon^2 + \rho^2}$ (Sect. 2), are indicated as probable members in Table 2. The star GCS\,0916200216 is too red and was classified as a non-member. 

Using the total-to-selective extinction ratio $R_V = 3.1$ we obtain $A_V = 0.12$. The absolute distance modulus is $(m-M)_{\circ}$ = 8.25  ${\pm}$ 0.20. The distance from the Sun is d$_{\odot}$ ${\approx}$ 0.45 ${\pm}$ 0.04 kpc, thus somewhat larger than that derived by SP. Assuming a solar Galactocentric distance of 8 kpc (Reid 1993), the Galactocentric coordinates are $X = -7.87$ kpc, $Y = -0.37$ kpc, $Z = -0.23$ kpc (X $<$ 0 is our side of the Galaxy). The Galactocentric distance is $R_{GC} = 7.88$ kpc, thus a nearly solar circle object. It is located at ${\approx}$ 230 pc below the Galactic plane, where younger disc components are less probable.

Astrometry in the region is significant for proper motions but not for parallaxes. In the CCD area HD\,35183 (HIP\,24652 or SP\,7) has a parallax p = 1.22 ${\pm}$ 0.87 mas and a corresponding distance d$_{\odot}$ = $820^{-342}_{+2037}$ pc. Within uncertainties this TO star of spectral type A2 has a distance compatible with that of NGC\,1901 derived from the CMD. The other Hipparcos stars are outside the CCD area. HD\,35230 (HIP\,24671) has p =  3.47 ${\pm}$ 0.78 mas and  d$_{\odot}$ = $288^{-53}_{+81}$ pc which is marginally compatible with that of NGC\,1901, but Johnson photometry V = 7.57, (B-V) = 0.86 provided in the Hipparcos catalogue, and most of all Tycho proper motions ${\mu_\alpha}$ = -17.4${\pm}$ 1.1 mas/yr, ${\mu_\delta}$ = -0.8${\pm}$ 1.1 mas/yr indicate that it is not a member. HD\,269320 (HIP\,24763) with p = 8.91 ${\pm}$ 1.97 mas and d$_{\odot}$ = $112^{-20}_{+32}$ pc is in the foreground. We show in Fig. 4 a sky chart for  a Tycho extraction of diameter 60$^{\prime}$ centered in NGC\,1901. The large circle encompasses the CCD data, where most of the stars share motions (Table 2) indicating a physical system, in agreement with Murray et al. (1969). The stars in  the NW extension basically move in the same direction as the central stars. The proper motion moduli are comparable: 9.3 ${\pm}$ 3.3 mas/yr for 10 stars in the central part and 12.4 ${\pm}$ 9.7 mas/yr for 4 stars in the NW extension. The dispersions are not intrinsic because they are of the order of the proper motion errors. A few additional stars are compatible with the central concentration and NW extension stars, and could be related to a corona - see M\,73 (Bassino et al. 2000). The age and distance of the stars in the central and NW concentrations are similar, and it remains to be studied whether the substructures are the result of the evolution of a parent initial cluster, an interaction or a close approach. In particular, projection effects of tidal arms might cause a secondary concentration, according to N-body simulations (Combes et al. 1999).

\begin{table*}
\caption[]{CCD Photometric Results for NGC\,1901}
\label{tab1}
\renewcommand{\tabcolsep}{0.70mm}
\renewcommand{\arraystretch}{0.5}
\begin{tabular}{lccccccccc}
\hline\hline
 Names$^{\mathrm{a}}$  &$\alpha$ & $\delta$& ~V ~&(B-V) & ~V ~&(B-V) &${\mu}_{\alpha}$&${\mu}_{\delta}$& Membership$^{\mathrm{b}}$\\
& h:m:s &$^{\circ}$:$^{\prime}$~~:$^{\prime\prime}$ & SP & SP & CCD & CCD & mas/yr & mas/yr &\\\hline
SP\,7, TCY2\,552 & 05:17:23.0 & -68:28:19 & 9.16 & 0.14 & 9.14 & 0.13 & 0.0 & 10.7 & pm \\
SP\,11, TYC2\,732  & 05:18:02.1 & -68:21:19 & 9.25 & 0.20 & 9.27 & 0.16 & -0.8 & 12.4 &  pm \\
SP\,14, TYC2\,142 & 05:18:22.5 & -68:28:02 & 10.24 & 0.21 & 10.23 & 0.22 & -0.7 & 11.3 & pm\\
SP\,12, TYC2\,669 & 05:18:11.9 & -68:25:36 & 10.38 & 0.24 & 10.40 & 0.24 & 4.6 & 6.8 & pm\\
SP\,10, TYC2\,883 & 05:17:59.2 & -68:31:28 & 10.48 & 0.25 & 10.44 & 0.23 & 3.6 & 6.6 & pm \\
SP\,9, TCY2\,498 & 05:17:35.8 & -68:21:47 & 10.91 & 0.31 & 10.99 & 0.30 & 5.0 & 11.1 & pm\\
SP\,13, TCY2\,702 & 05:18:16.7 & -68:25:10 & 11.37 & 0.45 & 11.34 & 0.44 & 5.0 & 5.5 & pm\\
SP\,15, TCY2\,196 & 05:18:29.5 & -68:27:14 & 11.58 & 0.49 & 11.55 & 0.48 & 3.0 & 8.9 & pm\\
SP\,16, TCY2\,590 & 05:18:42.7 & -68:27:33 & 11.64 & 0.41 & 11.68 & 0.43 & 1.1 & 8.3 & pm\\
SP\,8, GSC\,0916201005 & 05:17:27.1 & -68:25:42 & 12.56 & 0.58 & 12.78 & 0.56 & - & - & pm\\
GSC\,0916200682 & 05:18:41.8 & -68:22:29 & - & - & 12.25 & 0.72 & - & - & pm\\
GSC\,0916200834 & 05:19:05.3 & -68:30:46 & - & - & 12.78 & 0.62 & - & - & pm\\
GSC\,0916200626 & 05:18:19.1 & -68:26:25 & - & - & 12.9 & 0.56 & - & - & pm\\
GSC\,0916200464 & 05:18:27.7 & -68:31:28 & - & - & 13.16 & 0.23 & - & - & pLMC\\
GSC\,0916200216 & 05:17:51.2 & -68:24:34 & - & - & 13.52 & 1.12 & - & - & nm\\
MACS\,0517684015& 05:17:41.3 & -68:29:04 & - & - & 13.60 & 0.75 & - & - & pm\\
GSC\,0916200883 & 05:17:59.2 & -68:31:28 & - & - & 13.68 & 0.40 & - & - & pLMC\\
1 & 05:18:04.0 & -68:26:44 & - & - & 13.85 & 0.85 & - & - & pm\\
USNO\,015002864868 & 05:17:51.3 & -68:26:08 & - & - & 13.95 & 0.75 & - & - & pm\\
GSC\,0916200619 & 05:18:52.7 & -68:32:35 & - & - & 14.50 & 0.87 & - & - & pm\\
2 & 05:17:27.5 & -68:29:45 & - & - & 14.52 & 0.93 & - & - & pm\\
3 & 05:18:36.5 & -68:29:10 & - & - & 14.52 & 1.09 & - & - & pm\\
USNO\,01502856433 &  05:17:32.6 & -68:23:52 & - & - & 14.53 & 0.59 & - & - & pLMC\\
MACS\,0517684021& 05:17:56.5 & -68:24:22 & - & - & 14.71 & 0.70 & - & - & pLMC\\
USNO\,015002855788 & 05:17:31.1 & -68:23:51 & - & - & 14.99 & 0.68 & - & - & pLMC\\
MACS\,0518684007 & 05:18:11.4 & -68:26:34 & - & -  & 15.01 & 0.34 & - & - & pLMC\\
USNO\,015002858752&05:17:37.6 & -68:31:51 & - & - & 15.22 & 0.69 & - &- & pLMC\\
MACS\,0518685008 & 05:18:33.4 & -68:30:42 & - & -  & 15.24 & 0.34 & - & -& pLMC\\
MACS\,0518684004 & 05:18:09.4 & -68:26:26 & - & - & 15.63 & 0.36 & - &- & pLMC\\
4 & 05:18:58.0 & -68:21:25 & - & - & 16.12 & 0.85 & - & - & pLMC\\

\hline\hline

\end{tabular}

\begin{list}{}
\item {$^{\mathrm{a}}$\textit{SP} is from Sanduleak \& Philip (1968), \textit{GSC} Guide Star Catalogue, \textit{TCY2} part of the Tycho identifier, \textit{MACS} Tucholke et al. (1996), \textit{USNO} Monet et al. (1998); $^{\mathrm{b}}$\textit{pm} means probable, \textit{nm} non-member, \textit{pLMC} probable LMC member.} 
\end{list}
\end{table*}

\begin{figure} 
\resizebox{\hsize}{!}{\includegraphics{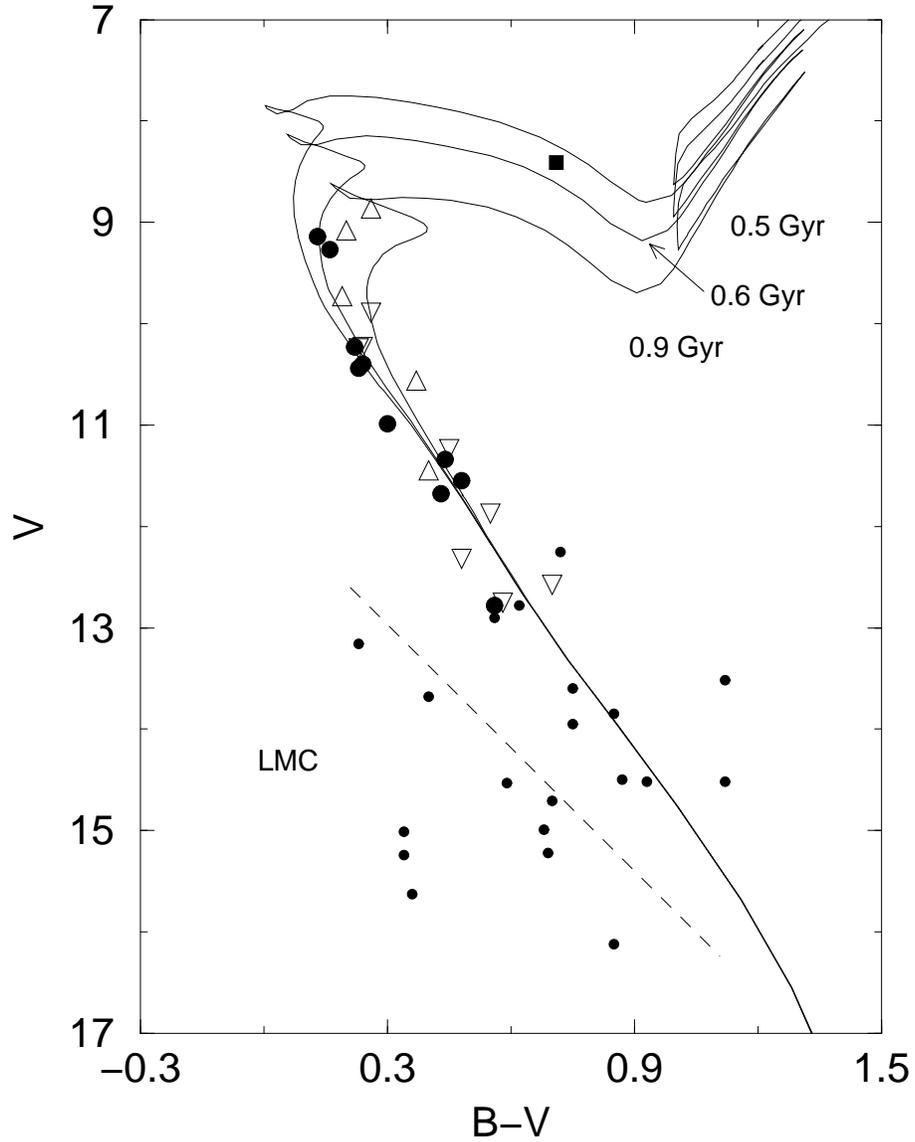}}
\caption[]{NGC\,1901: V$\times$(B-V) diagram for 30 stars with CCD photometry: small solid circles are new stars and large ones are in common with SP. Filled square is HD\,35294 with Tycho photometry. Open triangles are 13 SP stars outside the CCD frames: upwards ones  are in the NW extension and downwards ones are south of the CCD frames. Padova isochrone solution is shown. Dashed line separates probable LMC members.}
\label{fig1}
\end{figure}

\begin{figure} 
\resizebox{\hsize}{!}{\includegraphics{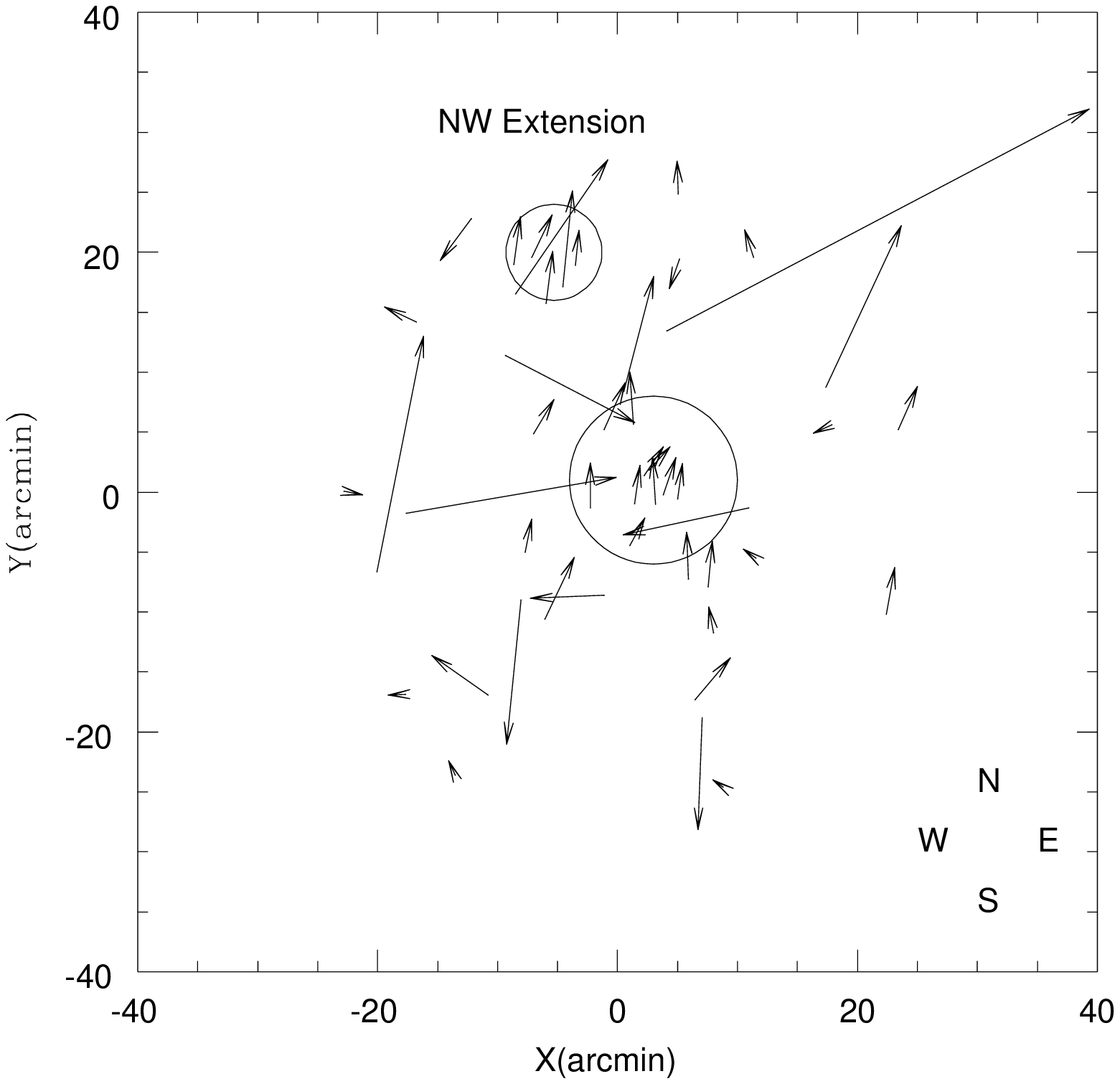}}
\caption[]{NGC\,1901: sky chart of Tycho stars in a region of diameter 60$^{\prime}$. Arrow length is proportional to  proper motion modulus, and largest one is  113.5 mas/yr. The large circle has a diameter of 12$^{\prime}$ and corresponds to the CCD area.}
\label{fig1}
\end{figure}

\section{The stellar group NGC\,1252}

Fig. 5 shows the V ${\times}$ (B-V) CMD with the 12 stars from the present photometry and total colour error bars $\sqrt{\epsilon^2 + \rho^2}$ (Sect. 2). Also shown are the 4 complementary  BT stars in the area (BT\,1, 11, 13 and 28). The fainter stars in the CMD were observed for the first time. The CMD suggests a MS and possibly some related giants. We used Padova isochrones of different ages, together with reddening constraints to find best fit solutions. 

\begin{figure}
\resizebox{\hsize}{!}{\includegraphics{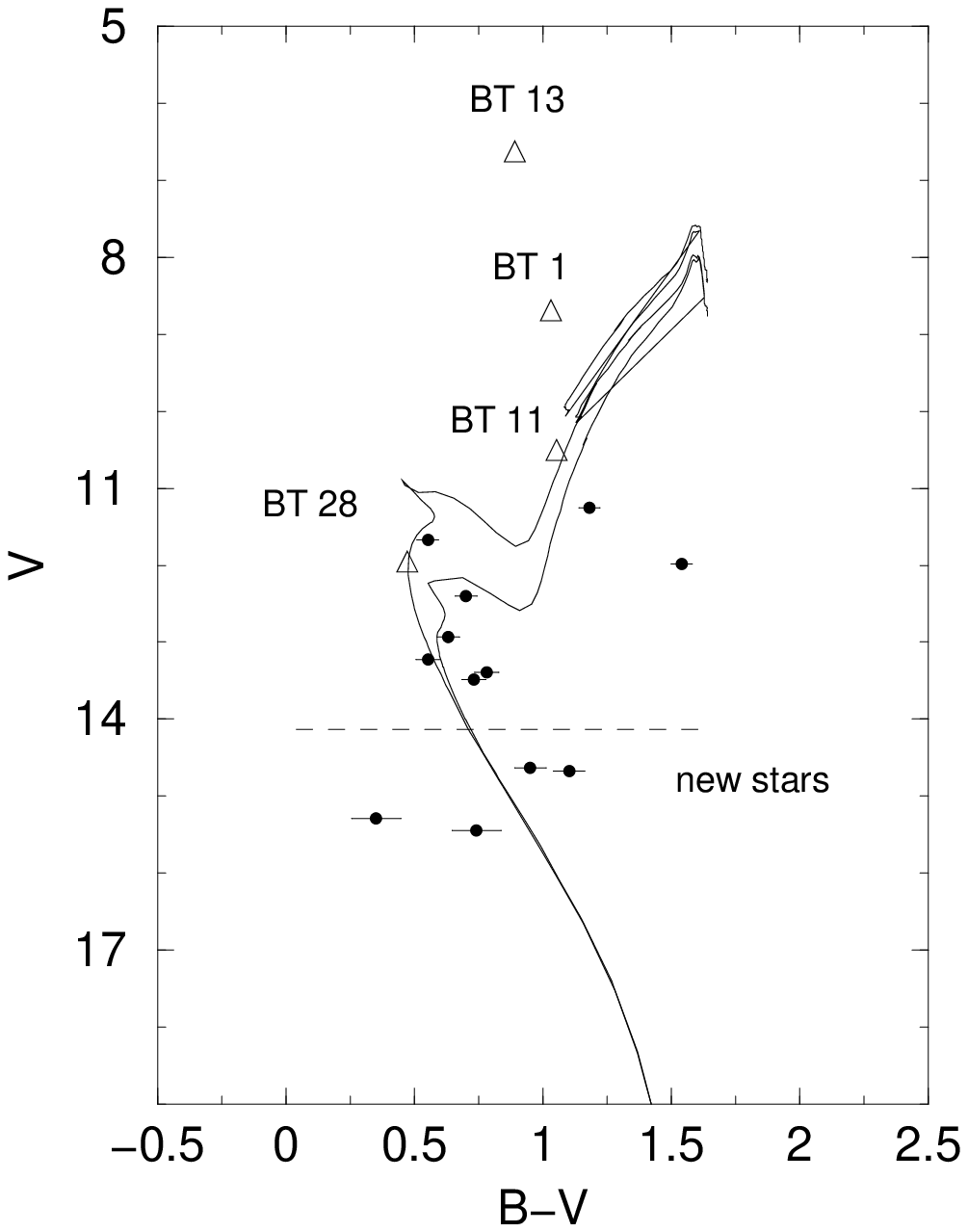}}
\caption[]{NGC\,1252: V$\times$(B-V) diagram for 12 stars with CCD photometry (solid circles). Total errors in colours are shown. Open triangles are stars from Bouchet \& Th\'e$^{\prime}$s (1983) photometry. Line separates stars observed for the first time. Padova isochrone solutions are shown.}
\label{fig1}
\end{figure}

The bright giants BT\,1 (K1 III/IV) and BT\,13 (G8 III) have Hipparcos parallaxes  p = 10.24 ${\pm}$ 0.78 mas and p = 5.48 ${\pm}$ 0.54 mas, respectively. The corresponding distances are d$_{\odot}$ = 98 ${\pm}$ 8 pc and d$_{\odot}$ = 182 ${\pm}$ 16 pc. Both stars together cannot belong to a possible physical stellar group. In Fig. 5 we superimpose the best isochrone fits. The reddening was estimated taking into account two methods. Schlegel et al.$^{\prime}$s (1998) reddening maps are based on dust thermal emission and provide in this direction $E(B-V) = 0.02$. Burstein \& Heiles$^{\prime}$ (1982) reddening maps are based on H\,I column densities and galaxy counts and indicate an absence of reddening in the area. We conclude  that $ E(B-V) = 0.02$ is compatible with the MS for intermediate ages and an apparent distance modulus $(m-M) = 9.1$. From the 2 and 4 Gyr isochrones we estimate an age 3${\pm}$1 Gyr for NGC\,1252. With $A_V = 0.062$ and $(m-M)_{\circ} = 9.04$ ${\pm} 0.25$ we obtain d$_{\odot}$ ${\approx}$ 0.64 ${\pm}$ 0.07 kpc. The distance modulus error is by far dominated by vertical uncertainties in the CMD fit. The Galactocentric coordinates are $X = -7.82$ kpc, $Y = -0.41$ kpc and $Z = -0.46$ kpc. The Galactocentric distance is $R_{GC}$ = 7.84 kpc, thus likewise NGC\,1901, NGC\,1252 is a nearly solar circle object. It is located at ${\approx}$ 460 pc below the Galactic plane, consistent with the old disc distribution (Friel 1995).

The star BT\,17 was classified as a non-member (Table 1) because it is too red with respect to the isochrones. BT considered the stars BT\,12, 14, 15, 16, 18, 19 and  27 as Galactic field of their large region, but their probable members were around TW Hor (Sect. 1). We considered them as probable members based on their locus with respect to the isochrone MS, except BT\,15 which is rather too red with respect to the subgiant branch and was classified as a less probable member. The CCD photometry of the Guide Star Catalogue entries  GSC 0849800928, 0849800945 and 0849801024 (Table 1) allowed us to classify them as probable members taking into account the colour error bars and the possibility of a double star sequence parallel to the MS.  The star GSC\,0849801321 is classified as a non member because it is much bluer than the isochrones$^{\prime}$ MS. Indeed it is located in the CMD region corresponding to the bulk of field stars (Sect. 4.1). The stars BT\,1 and 13 are bluer than the isochrone giant branch, which indicates that they are not members. This is supported by their large parallaxes, locating them closer than what we find for NGC\,1252. BT\,11 is a probable member from its subgiant branch locus. Finally, BT\,28 appears to be a probable member belonging to the TO region.

The present gain of ${\Delta}$V ${\approx}$ 2 mag with respect to BT allows one to probe lower along the MS , which is a constraint for the TO, and isochrone fitting. Three additional GSC stars are present in the CCD images. According to the GSC, they  are slightly fainter than  those measured in Table 1. Assuming the additional stars as members their total number in the range  14 $< V <$ 16 would be 6. Note that the isochrones (Fig. 5) have most probable members on the red MS side. This suggests them preferentially as part of a double star sequence, which is  expected for a cluster remnant stellar content. Binary star sequences have been observed e.g. in the globular cluster 47 Tucanae (Santiago et al. 1996a) and the young LMC cluster NGC1818 (Elson et al. 1998). Globular clusters appear to have a smaller binary fraction than open clusters. In the globular cluster NGC\,2808 Ferraro et al (1997) obtained a fraction of 24 \%. For the post-core-collapse globular cluster NGC\,6752 (Rubenstein \& Bailyn 1997) the overall fraction is low ($<$ 16 \%) except in the inner core (15 - 38 \%). Gonz\'alez \& Lapasset (2000) estimated  a fraction of 26 \% (46 \% including all suspected binaries) in the open cluster NGC\,2516. From 167 bright Hyades members (Patience et al. 1998) the detected multiplicity is 59 binaries and 10 triples. This dynamically evolved open cluster has a detected binary fraction of 41 \%. Binary fractions in globular clusters cannot be directly compared to those in open clusters, since loose binaries are thought to be destroyed at initially higher densities. Nevertheless in an advanced dynamical state for an open cluster or remnant one does not expect binaries to be destroyed by stellar interactions. Comparing the number of MS and TO stars on either side of the fitted isochrones for NGC\,1252 (Fig. 5) a high binary fraction of 67 \% is obtained. The statistics is low, but taking this fraction at face value NGC\,1252 would be consistent with a late dynamical state.

\subsection{Comparison with a field model}

We wish to compare the concentration CMD with that expected from field
stars in the same direction and limiting magnitude, following Bassino et al. (2000). Bica et al. (2001) have compared
integral star counts down to a limiting magnitude in their fields to 
those expected from a Galactic structure model. We
use the same model as starting point to build 
a theoretical CMD of field
stars in the direction of NGC 1252.
This model has been described in more detail by Reid \& Majewski (1993)
and Santiago et al. (1996b). A recent application of the method to Galactic fields 
was carried out by Kerber et al. (2001). It includes 3 structural components: a
thin disc, a thick disc and spheroid. 
The discs stellar density profiles are described by double exponentials,
one along the plane of the disc and the other perpendicular to it. The stellar luminosity function is well
constrained down to M$_V$ ${\approx}$ 12. 
Given the magnitude limit (V ${\approx}$ 15.5) for the concentration stars,
our star counts are almost entirely dominated by the thin disc.

Given a direction, limiting magnitude and solid angle, the model provides 
expected numbers of stars throughout the CMD plane.
In Fig. 6 we show a model CMD for a region of diameter 1$^{\circ}$ around
the concentration.
This large area provides a  statistically significant CMD.
The isochrones on the right panel are from Fig. 5.
They do not fit the bulk of field stars, which are considerably
bluer in average. The isochrones are an upper envelope to the bulk of field stars and consequently NGC\,1252 appears to be located in the field foreground. 
In Fig. 7 we present a similar CMD but for the small solid angle of the CCD
data. This sample was randomly 
taken from the expected model counts in each CMD
cell (scaled to the smaller solid angle) and allowing for Poisson fluctuations. As expected, a typical small field CMD will be populated preferentially 
with stars from the densest CMD zone in the large area (Fig. 6). However, although  statiscally unlikely, one cannot rule out a fluctuation producing
a CMD similar to that of the concentration.
The model CMD (Fig. 6) extends down to V ${\approx}$ 17, since unlike the CCD data limit (Fig. 7), it is not affected
by incompleteness. In the particular small field sampling the isochrones are located redwards of the artificial stars sampled. 
The NGC\,1252 stars suggesting a MS (Fig. 5)
are distributed in a region much narrower than the model stars 
(Fig. 6). These comparisons give support to the existence of a physical stellar group.

\begin{figure}
\resizebox{\hsize}{!}{\includegraphics{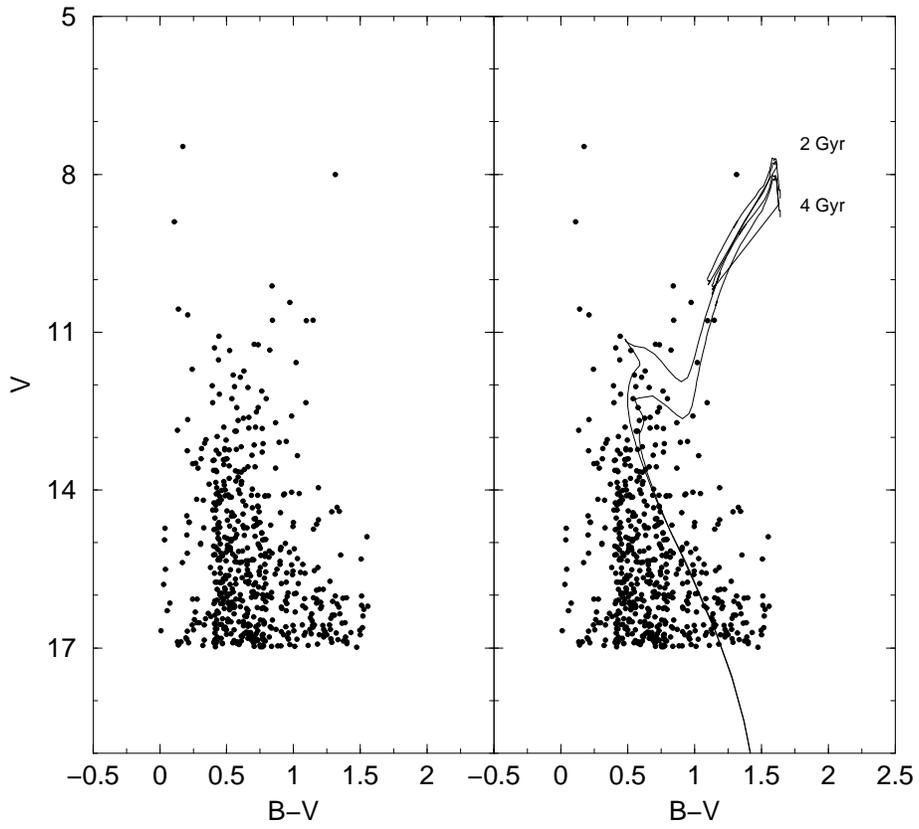}}
\caption[]{Left panel: V ${\times}$ (B-V) diagram predicted
by the Galactic structure model for a region of diameter 1$^{\circ}$ in the same direction as NGC\,1252.
Right panel: same isochrone solution for NGC\,1252 as in Fig. 5.}
\label{fig1}
\end{figure}

\begin{figure} 
\resizebox{\hsize}{!}{\includegraphics{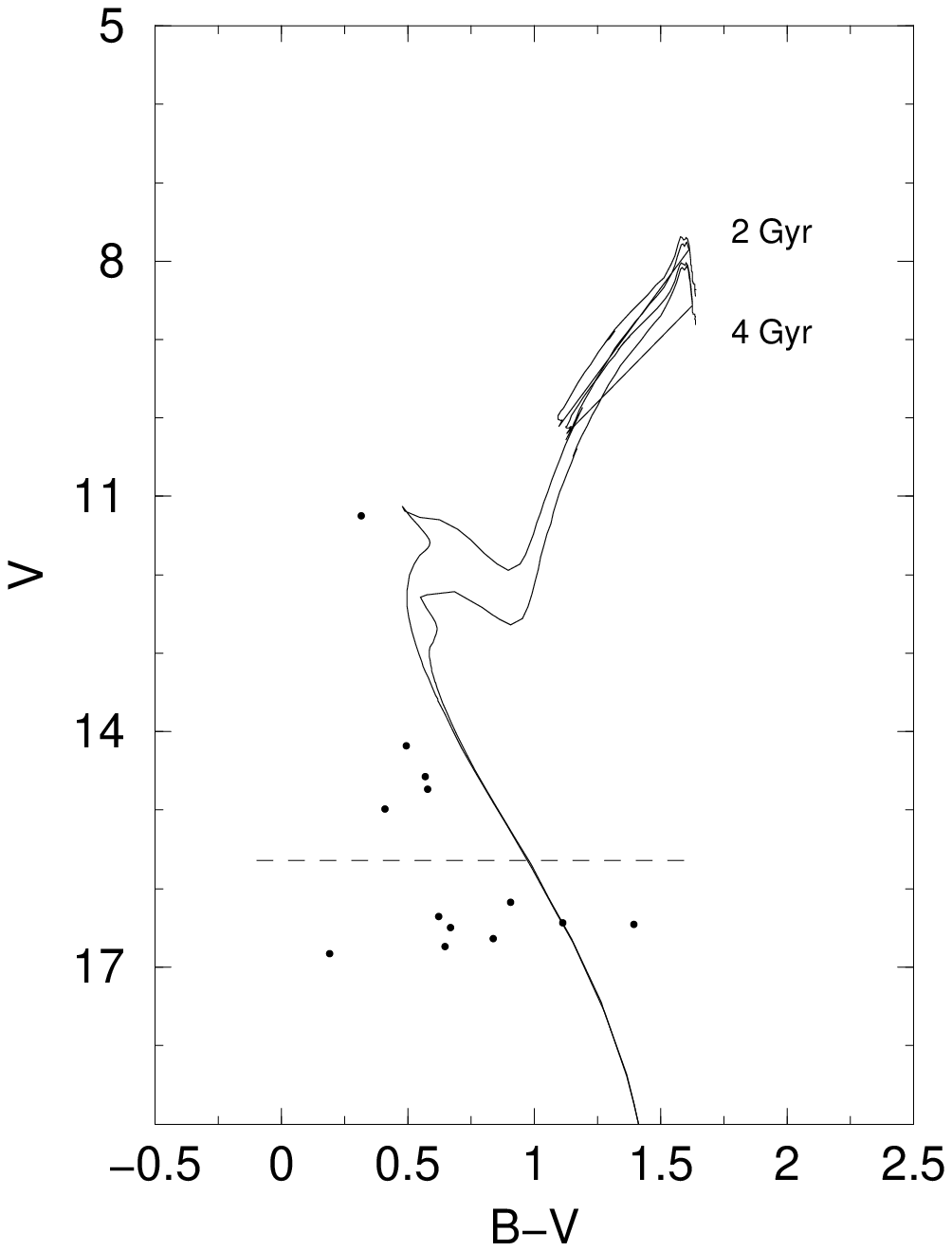}}
\caption[]{Theoretical V ${\times}$ (B-V) Galactic field diagram for a solid angle equal to that of the CCD. Same isochrone solution for NGC\,1252 (Fig. 5). Dashed line indicates the CCD limit.}
\label{fig1}
\end{figure}

\subsection{Available astrometry}

We extracted Tycho stars within a 80$^{\prime}$ diameter centered in the concentration. We show in Fig. 8 a sky chart with  proper motions proportional to arrow sizes. The large circle has a diameter of 14$^{\prime}$ and corresponds to the present stellar concentration. Note that stars fainter than V ${\approx}$ 12 are absent in the Tycho catalogue. The bright giants BT\,1 and 13 have different proper motions, in agreement with their different parallaxes. They are foreground field giants with respect to NGC\,1252. The probable member BT\,11 and the less probable one BT\,15 (Table 1) present proper motions significantly different (${\mu_\alpha}$ = +15.7${\pm}$ 4.6 mas/yr, ${\mu_\delta}$ = +17.7${\pm}$ 4.2 mas/yr and ${\mu_\alpha}$ =- 0.4 ${\pm}$ 5.4 mas/yr, ${\mu_\delta}$ = +2.2 ${\pm}$ 4.9 mas/yr, respectively), and they cannot be members simultaneously. Outside the concentration area (Fig. 8) three Hipparcos stars occur: HIP\,14930 (TW Hor) at d$_{\odot} = 403_{-74}^{+118}$ pc, HIP\,15176 at d$_{\odot}$ = $397_{-149}^{+603}$ pc and HIP\,14975 at d$_{\odot}$ = $630_{-210}^{+630}$ pc. From the distances and/or Johnson photometry in the Hipparcos catalogue, we concluded that they are non-members.

\begin{figure}
\resizebox{\hsize}{!}{\includegraphics{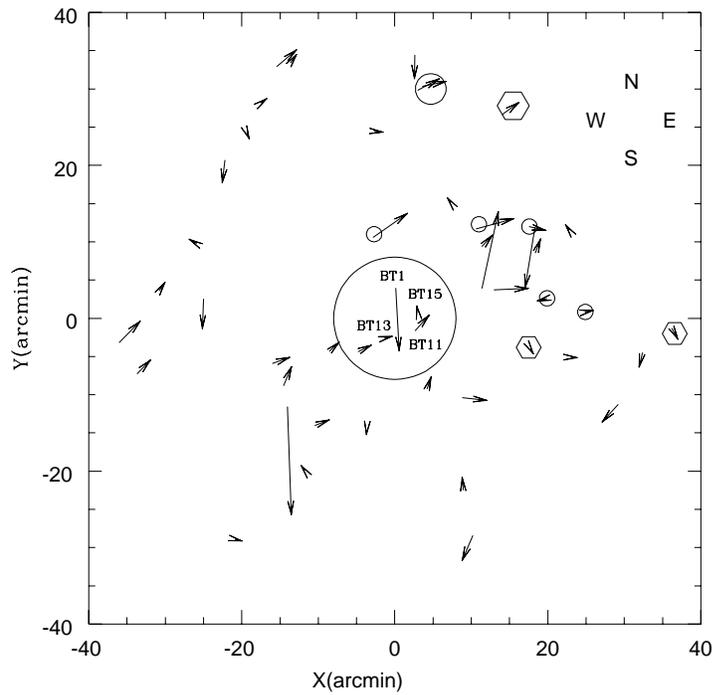}}
\caption[]{NGC\,1252: sky chart of Tycho stars in a region of diameter 80$^{\prime}$ centered in the present concentration. Arrow length is proportional to  proper motion modulus, the largest one is 121.2 mas/yr. Large circle has a diameter of 14$^{\prime}$ and encompasses NGC\,1252. Hexagones are Hipparcos stars outside it. Small circles are stars with proper motions and photometry compatible within uncertainties with NGC\,1252.}
\label{fig1}
\end{figure}

The available proper motions are not deep enough to study the TO and MS (Fig. 5). The faint stars (12 $< V <$ 15.5) would be necessary for a conclusive proper motions result. A cautionary remark for proper motions and parallaxes based on observations spanning a few years, such as Tycho and Hipparcos data, is binarity. Early plates in the XX century like ACT$^{\prime}$s combined to subsequent images can lessen the problem, but not for binaries with periods longer than 100-200 years (Baumgardt 1998).

Based on N-body simulations Terlevich (1987) showed that a corona is expected after 300-400 Myr of evolution, and that such stars are likely to escape in the near future. We show in Fig. 9 the CMD of stars {\it outside} the concentration area with BT photometry. We distinguish BT$^{\prime}$s members and non-members. We  superimpose the present isochrone solution (Fig. 5), which  suggests that most of their \textit{ non-members} in the TO region are compatible with our interpretation of NGC\,1252. A few giants considered by them to be members are as well possible members in our interpretation owing to their proximity to the isochrones. We conclude that 13 stars are compatible with the isochrones. Eight are marked in the Tycho chart (Fig. 8), since 2 are absent in the Tycho catalogue and 3 are outside the extraction. The marked stars are distributed on one side of the concentration because it is located at the edge of BT$^{\prime}$s large region. Six marked stars have proper motions compatible within uncertainties either with the probable member BT\,11 or the less probable member BT\,15, and they could be members of a corona, as seen in M\,73 (Bassino et al. 2000). Tycho errors can be as large as 5 mas/yr for these stars, which at the object distance of 0.64 kpc implies transversal velocity errors of ${\approx}$ 15 km/s. Since velocity dispersions in open clusters are much lower (e.g. 0.23 km/s for the Hyades, Gunn et al. 1988) no inference can be made about internal motions in the present object.

\begin{figure}
\resizebox{\hsize}{!}{\includegraphics{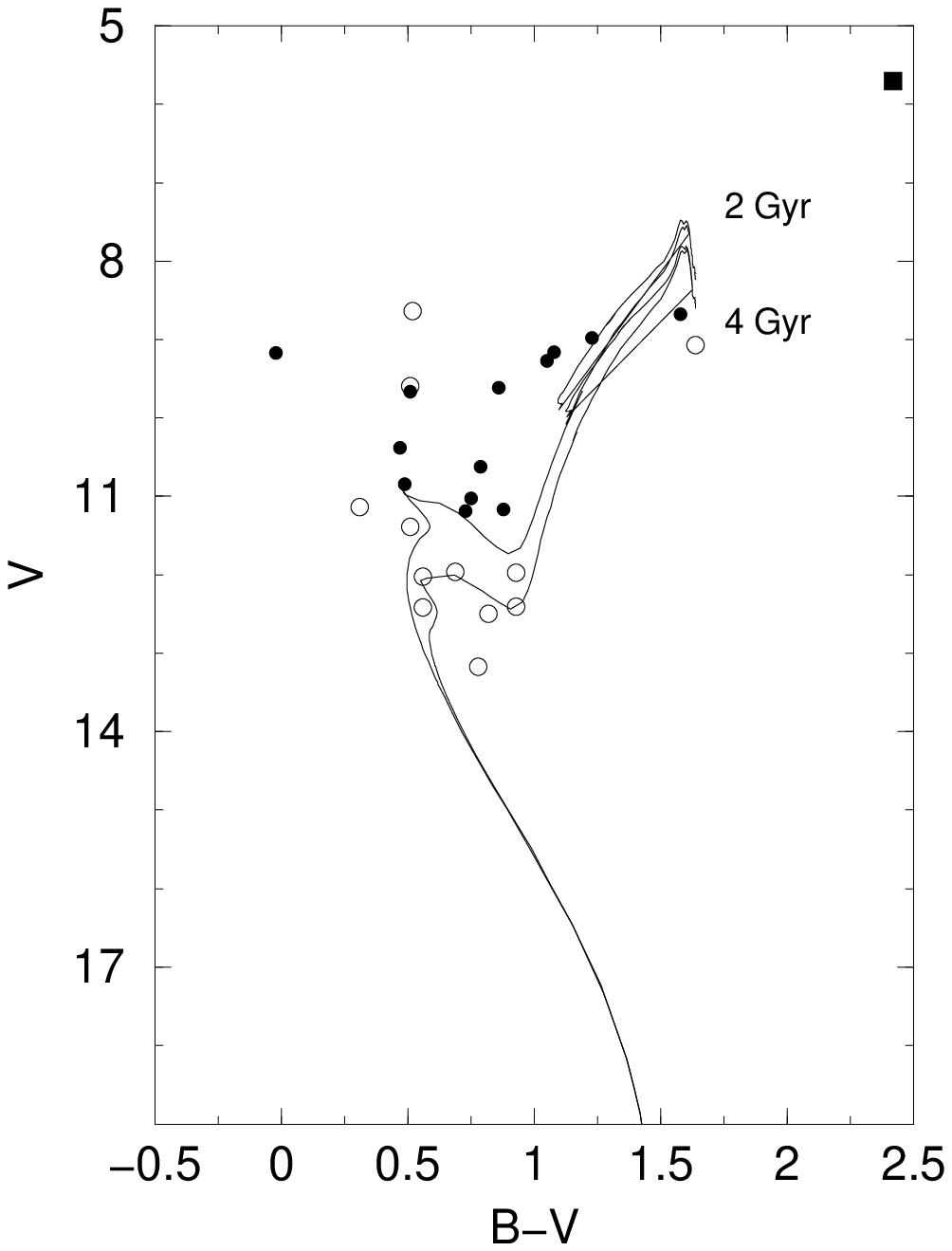}}
\caption[]{V ${\times}$ (B-V) diagram for stars in BT$^{\prime}$s photometry located outside the concentration area. Solid circles are BT$^{\prime}$s member stars and open circles non-member stars according to their interpretation. Continuous lines: same NGC\,1252  isochrone solution as in Fig. 5.}
\label{fig1}
\end{figure}

\section{Comparisons with the Hyades}

The luminosity function (LF) can provide information on cluster mass and dynamical state. We take the Hyades as template  to infer the properties of the present objects. We have selected and homogenized stars with V $<$ 11 (WEBDA), which encompass the available absolute mag intervals for  NGC\,1901 and NGC\,1252. Adopting (m-M)$_0$ = 3.32 for the Hyades (WEBDA, Weidemann et al. 1992) we computed M$_V$ for corresponding giants, TO and MS stars which form the reference histograms in Fig(s). 10 and 11. For NGC 1901 and NGC 1252 we computed the M$_V$ histograms using the derived parameters and their 20 and 12 probable members, respectively (Sects. 3 and 4). We employed 181 Hyades members in the magnitude range of Fig. 10, and 147 in that of Fig. 11.

We show in Fig. 10 percentage fractions with respect to the total population of probable members in the available absolute mag range for NGC 1901, as compared to the Hyades. The histogram shapes are similar indicating a comparable mass function  slope in the available range. CCD observations in the whole object area would improve the statistics in the present mag range, and  deeper observations might reveal the behaviour of the low MS. The population ratio of stars in the two objects is R$_P$ = 9.1, which suggests that the Hyades are this factor more massive than NGC\,1901 as sampled by the CCD. However, SP members with proper motions from Murray et al. (1969) and the present Tycho analyses indicate 10 bright members in the CCD area and 10 outside it. This would indicate a mass ratiom ${\cal{M}}_{Hyades}$/ ${\cal{M}}_{NGC\,1901}$ = 4.6, assuming that both objects are in a similar dynamical state with equally populated lower MS.

In Fig. 11 we compare the LFs of NGC\,1252  and the Hyades. The shapes are similar despite the limited population of stars for NGC\,1252. In this case the object is essentially contained in the CCD so that R$_P$ = 12.3, and the mass ratio is thus ${\cal{M}}_{Hyades}$/${\cal{M}}_{NGC\,1252}$ = 12.3, again under the assumption  that both objects are in a similar dynamical state. It would be important to probe the lower MS (M$_V >$ 6).

The Hyades have a  nearly flat current mass function and appear to have lost ${\approx}$ 90 \% of an estimated total \textit{initial} mass of 14000 M $_{\odot}$ (Weidemann et al. 1992). In this sense the Hyades can be considered an open cluster remnant despite the fact that they are widely accepted as an open cluster. NGC\,1901 and NGC\,1252 have similar LFs as that of the Hyades in the available luminosity range, although they differ in terms of scaling factor. The inferred lower total  masses, especially that of NGC\,1252 for which  evidence of a high binary fraction was found (Sect. 4), favour the concept of an open cluster remnant.

\begin{figure}
\resizebox{\hsize}{!}{\includegraphics{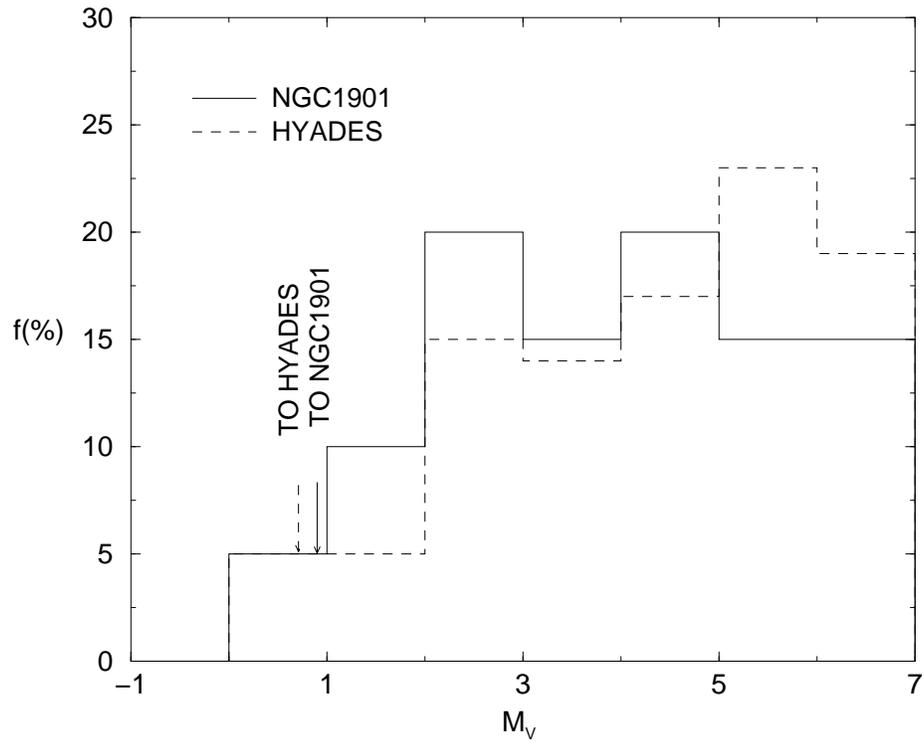}}
\caption[]{Absolute V luminosity function of NGC\,1901 compared to that of the Hyades. Percentage fractions f(\%) with respect to the total population of probable members in the magnitude range are shown.}
\label{fig1}
\label{fig1}
\end{figure}

\begin{figure}
\resizebox{\hsize}{!}{\includegraphics{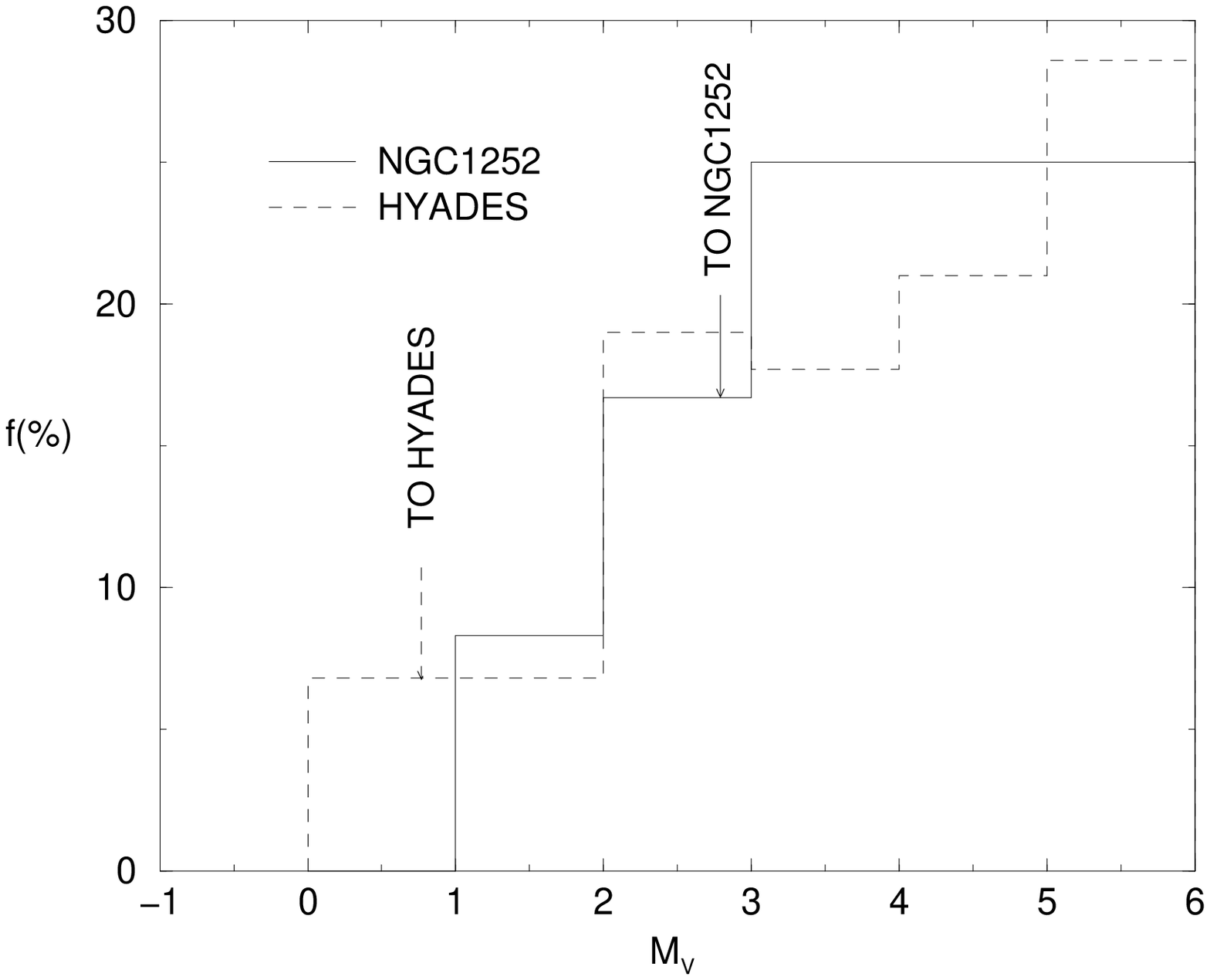}}
\caption[]{Same as Fig. 10, for NGC\,1252.}
\label{fig1}
\end{figure}

\section{Concluding remarks}

We provide CCD photometry of the stellar groups  NGC 1901 and NGC\,1252, probing them deeper than in the literature. We also explore the available Hipparcos and Tycho astrometries. The photometry confirms that NGC\,1901 is a physical system which is also supported by proper motions. For NGC\,1252 we describe a different set of stars as compared to the object interpreted by Bouchet \& Th\'e (1983), and questioned by Eggen (1984) and Baumgardt (1998). The present CMD suggests a TO and MS, so that a physical stellar group cannot be ruled out. The available astrometry is not helpful enough to constrain properties of this object. 

NGC\,1901 has an age comparable to that of the Hyades and their dynamical states are similar within the available LF range for NGC\,1901. It would be important to further deepen NGC\,1901$^{\prime}$s photometry to verify if depletion of lower MS stars occurs with respect to the Hyades, which would be the signature of a more evolved dynamical state. NGC\,1252 is consistent with the notion of an old open cluster remnant, similar to M\,73 (Bassino et al. 2000). NGC\,1901 appear to a factor 4.6 less massive than the Hyades, while NGC\,1252 is a factor 12.3, assuming equally populated lower main sequences.

In order to conclusively determine the nature and dynamical state of stellar groups like NGC\,1901 and NGC 1252 one needs deeper photometry and astrometry, and high dispersion spectroscopy for radial velocities and abundances. Binary fraction studies are fundamental too. Finally, from theoretical and numerical calculations it would be important to establish criteria to classify late stages of open cluster dynamical evolution and subsequent remnants. The inclusion of binaries and triples is crucial in the models. As a first step, the Hyades themselves require a clear determination of the dynamical state and classification as open cluster or open cluster remnant.

\subsection*{Acknowledgements}
We thank an anonymous referee for important remarks, the students M. Campos and M. Aguero and the Bosque Alegre staff for assistance during the observations and hospitality, and M\'arcio R. de Oliveira (CNPq fellow) for discussions. We also thank Gerry Gilmore and Jim Lewis (IoA Cambridge) for the original software used for modeling the structure of the Galaxy. Part of the present data was extracted from the CDS database (Strasbourg). This work was partially supported by the Argentinian -Brazilian grant IRA 98/0967-9 by CONACyT/FAPERGS. We also acknowledge support from the Brazilian Institutions CNPq and CAPES.

\end{document}